\documentstyle[twocolumn,aps,epsfig]{revtex}

\begin{document}
\title{Overpopulation of $\overline \Omega$ in pp collisions:\\
a way to distinguish statistical hadronization from string dynamics}

\author{M. Bleicher\dag, F. M. Liu\dag, A. Ker\"anen$^1$,
J. Aichelin\dag, S.A. Bass\ddag, F. Becattini$^2$,
K. Redlich$^{3,4}$, K. Werner\dag}

\address{\dag\ SUBATECH,
Laboratoire de Physique Subatomique et des Technologies Associ\'ees \\
University of Nantes - IN2P3/CNRS - Ecole des Mines de Nantes \\
4 rue Alfred Kastler, F-44072 Nantes Cedex 03, France}

\address{\ddag\ Department of Physics, Duke University, Durham, NC, USA\\
    and RIKEN-BNL Research Center, Brookhaven National Laboratory,
    Upton, NY, USA}

\address{$^1$\   Department of Theoretical Physics, University of Oulu,
SF-90570 Oulu, Finland}

\address{$^2$\ Universit\`a di Firenze and INFN Sezione di
Firenze, Via G. Sansone 1, I-50019, Sesto F.no, Florence, Italy}
\address{$^3$\ Theoretical Physics  Division, CERN, CH-1211 Geneva 23, Switzerland}

\address{$^4$\ Institute for Theoretical Physics, University of
Wroclaw, PL-50204  Wroclaw, Poland}

\maketitle

\noindent
\begin{abstract}
The $\overline{\Omega}/\Omega$ ratio originating from string
decays is predicted to be larger than unity in proton proton
interactions at SPS energies ($E_{\rm lab}$=160 GeV). The
anti-omega dominance increases with decreasing beam energy. This
surprising behavior is caused by the combinatorics of
quark-antiquark production in small and low-mass  strings. Since
this behavior is not found in a statistical description of hadron
production in proton proton collisions, it may serve as a key
observable to probe the hadronization mechanism in such
collisions. \vspace{.6cm}
\end{abstract}

Hadron yields and their ratios stemming from the final state of
ultra-relativistic heavy-ion collisions have been extensively used
to explore the degree of chemical equilibrium
\cite{qgpreviews,stoecker,stock86a,rafelski,rafelski2,cleymans,johanna,braun-munzinger,spieles97a}
and to search for evidence for exotic states and phase transitions
in such collisions \cite{qgpreviews}. Under the assumption of
thermal and chemical equilibrium, fits with a statistical
(thermal) model have been used to extract bulk properties of hot
and dense matter, e.g. the temperature and chemical potential at
which chemical freeze-out occurs
\cite{rafelski2,cleymans,johanna,braun-munzinger,spieles97a}.

The application of a statistical model to elementary hadron hadron
reactions was first proposed by Hagedorn \cite{hag1} in order to
describe the exponential shape of the $m_t$-spectra of produced
particles in p+p collisions. Recent analysis \cite{becattini97a}
on hadron yields in electron positron and proton proton
interactions at several center-of-mass energies have shown that
particle abundances  can also be  described by a statistical
ensemble with maximized entropy. In fact, they are consistent with
a model assuming the existence of equilibrated fireballs at a
temperature $T \approx $160-170 MeV. These findings have given
renewed rise to the interpretation that hadronization in
elementary hadron hadron (hh) collisions is a statistical process,
which is difficult to reconcile with the popular picture that
hadron production in hh collisions is due to the decay of color
flux tubes \cite{colflux}, a model which has explained many
dynamical features of these collisions.

In this letter we argue  that  the $\overline{\Omega}/\Omega
\equiv \Omega^+/\Omega^-$ ratio in elementary proton proton
collisions is an unambiguous and sensitive probe to distinguish
particle production via the break up of a color flux tube from
statistical hadronization.

Color flux tubes, called strings, connect two SU(3) color charges
[ $3$ ] and [ $\overline 3$ ] with a linear confining potential.
If the excitation energy of the string is high enough it is
allowed to decay via the Schwinger mechanism \cite{schwinger},
i.e. the rate of newly produced quarks is given by:
\begin{equation}
\frac {{\rm d}N_{\kappa}}{{\rm d}p_{\perp}} \sim {\rm
exp}\left[-\pi m_{\perp}^2/\kappa \right],
\end{equation}
where $\kappa$ is the string tension and $m_{\perp} =
\sqrt{p_{\perp}^2 +m^2}$ is the transverse mass of the produced
quark with mass $m$.

However, specific string models may differ in their
philosophy and the types of strings that are created:
\begin{itemize}
\item
In UrQMD\cite{urqmdmodel} the projectile and target protons become
excited objects due to the momentum transfer in the interaction.
The resulting strings, with at most two strings being formed,  are
of the diquark quark type.
\item
In NeXuS\cite{nexusmodel}, the pp interaction is described in
terms of pomeron exchanges or ladder diagrams. Both hard and soft
interactions take place in parallel. Energy is shared equally
between all cut pomerons and the remnants. The endpoints of the
cut pomerons (i.e. the endpoints of the strings) may be valence
quarks, sea quarks or antiquarks.
\item
In PYTHIA\cite{pythiamodel}, a  scheme similar to that  in UrQMD
is employed. However, hard interactions may create additional
strings from scattered gluons and sea quarks. Most strings are
also of diquark quark form.
\end{itemize}

\begin{figure}[h]
\vskip 0mm \vspace{0cm}
\centerline{\psfig{figure=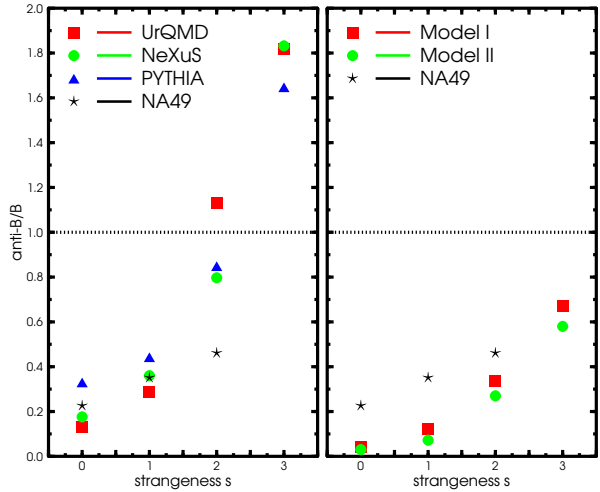,width=3.2in}} \caption{Left:
anti baryon to baryon ratio at $|y-y_{\rm cm}|<1$ in pp
interactions at 160 GeV as given by PYTHIA, NeXuS and UrQMD.
Right: anti baryon to baryon ratio for the same reaction as given
by statistical models. Stars depict preliminary NA49 data for the
$\overline B /B$ ratio at midrapidity. \label{pythia}}
\vspace{-0.4cm}
\end{figure}

Fig.~\ref{pythia} (left)
 depicts the anti baryon to baryon ratio at
midrapidity in proton proton interactions at 160 GeV. The results
of the   calculations by NeXuS, UrQMD and PYTHIA, which are well
established string-fragmentation models for elementary hadron
hadron interactions, are included in this figure. In all these
models, the $\overline B /B$ ratio increases strongly with the
strangeness content of the baryon. For strangeness $|s|=3$ the
ratio significantly exceeds unity. In UrQMD and PYTHIA the
hadronization of the diquark quark strings leads directly to the
overpopulation of $\overline \Omega$. In NeXuS, however, the
imbalance of quarks and anti quarks in the initial state leads to
the formation of $q_{\rm val}-\overline s_{\rm sea}$ strings (the
$s_{\rm val}-\overline q_{\rm sea}$ string is not possible). These
strings then result in the overpopulation of $\overline
\Omega^,$s.

In Fig.~\ref{pythia} the string model results are compared with
the predictions of statistical models (SM). Within the SM, hadron
production is commonly described using the grand canonical (GC)
partition function, where the charge conservation is controlled by
the related chemical potential. In this description a net value of
a given U(1) charge is conserved on average. However, in the limit
of small particle multiplicities, conservation laws must be
implemented exactly. This is done by using the canonical (C)
ensemble \cite{becattini97a,ko1,ahmed}. The conservation of
quantum numbers in the canonical approach severely reduces the
phase space available for particle production. Thus, exact charge
conservation is of crucial importance in the description of
particle yields in proton-induced processes and  in $e^+e^-$
\cite{becattini97a}, as well as in peripheral heavy-ion collisions
\cite{ahmed}.

In Fig.~\ref{pythia} the predictions of two different statistical
models for ${\overline B}/B$ ratios in pp collisions are included.
The main difference between these models are the implementation of
baryon number and electric charge conservation and the way an
additional strangeness suppression is introduced.

({\bf I}) The calculation in this statistical model
\cite{becalast} is a full canonical one with fixed baryon number,
strangeness and electric charge identical to those of the initial
state. Also, an extra strangeness suppression is introduced to
reproduce the experimental multiplicities. This is done by
considering the number of newly produced $\langle s \overline s
\rangle$ pairs as an additional charge to be found in the final
hadrons. The $s\overline s$ pairs fluctuate according to a Poisson
distribution and its mean number is considered as a free parameter
to be fitted \cite{becalast}. The parameters used for the
prediction of the $\Omega^+/\Omega^-$ ratio ($T$, global volume
$V$ sum of single cluster volumes and $\langle s \overline s
\rangle $) have been obtained by a fit to preliminary NA49 pp data
\cite{na49pp} yielding $T=183.7 \pm 6.7$ MeV, $VT^3 = 6.49 \pm
1.33$ and $\langle s \overline s \rangle = 0.405 \pm 0.026$ with a
$\chi^2/{\rm dof}= 11.7/9$. It must be pointed out that the
$\Omega^+/\Omega^-$ ratio is actually independent of the $\langle
s \overline s \rangle$ parameter and only depends on $T$ and $V$
(see also Fig.~\ref{om}).

({\bf II}) Here the conservation of baryon number and electric
charge is approximated by using the GC ensemble. Under thermal
conditions at top SPS energy this approximation leads to
deviations from the exact C results in pp collisions by at 
most 20$\%$-30$\%$ \cite{marek}. Strangeness conservation  is, however,
implemented on the canonical level following the procedure
proposed in \cite{ahmed}. It accounts for strong correlations of
produced strange particles due to constraints imposed by the
locality of the conservation laws. In pp collisions strangeness is
assumed not to be distributed in the whole volume of the fireball
but to be locally strongly correlated. A correlation volume
parameter $V_0=4\pi R_0^3/3$ is introduced, where $R_0\sim 1$ fm
is a typical scale of QCD interactions. The previous analysis of
WA97 pA data yields: $R_0\sim 1.12$~fm corresponding to $V_0\simeq
6$ fm$^3$. Note that, however, hidden strange particles are not
canonically suppressed in this approach. The analysis of
experimental data in AA collisions has shown that  $T$ and $\mu_B$
are almost entirely determined by the collision energy and only
depend weakly on the number of  participants \cite{cleymans}. The
$4\pi$ results of NA49 on $\bar p/\pi$ and $\pi /A_{part}$ ratios
in p-p and Pb-Pb collisions coincide within 20$\%$-30$\%$. In terms of
the SM this can be understood if $T$ and $\mu_B$ in p-p and Pb-Pb
collisions have similar values. We take $T\simeq 158$ MeV and
$\mu_B\simeq 238$ MeV as obtained from the SM analysis of full
phase-space NA49 Pb Pb data \cite{cleymans}. The volume of the
fireball $V\sim 17$ fm$^3$ and the charge chemical potential in pp
was then found to reproduce the average charge and baryon number
in the initial state.

The predictions of the statistical models are shown in
Fig.~\ref{pythia} (right). In these approaches the $\overline B
/B$ ratio is seen to exhibit a significantly weaker  increase with
the strangeness content of the baryon than expected in the string
fragmentation models. For comparison, both figures include
preliminary data on the $\overline B /B$ ratios obtained at
midrapidity by the NA49 Collaboration \cite{na49pp}.

\begin{table}
\begin{tabular}{llrr}
&Model    & $\Omega$ ($\times 10^{-4}$)& $\overline \Omega$
($\times 10^{-4}$)\\\hline
&NeXuS    & 0.48 & 0.79\\
&PYTHIA   & 0.17 & 0.30\\
&UrQMD    & 0.66 & 1.05\\
&Canonical Model {\bf I} & 0.46 & 0.31\\
&Canonical Model {\bf II}  &  0.41   &0.24 \\
\end{tabular}
\caption{\label{table1} Predictions of different models  on the
$4\pi$ yield of  $\Omega$ and $\overline \Omega$  in pp collisions
at 160 GeV.}
\end{table}
\noindent Note that the predictions of the statistical models in
Fig.~1 refer to full phase-space particle yields whilst
measurements of $\overline B /B$ ratios in pp collisions have been
performed at midrapidity, where they are expected to be the
largest. Therefore, sizeable deviations of the model results from
the data seen in Fig.~1 are to be expected.
\begin{figure}[t]
\vskip 0mm \vspace{0cm}
\centerline{\psfig{figure=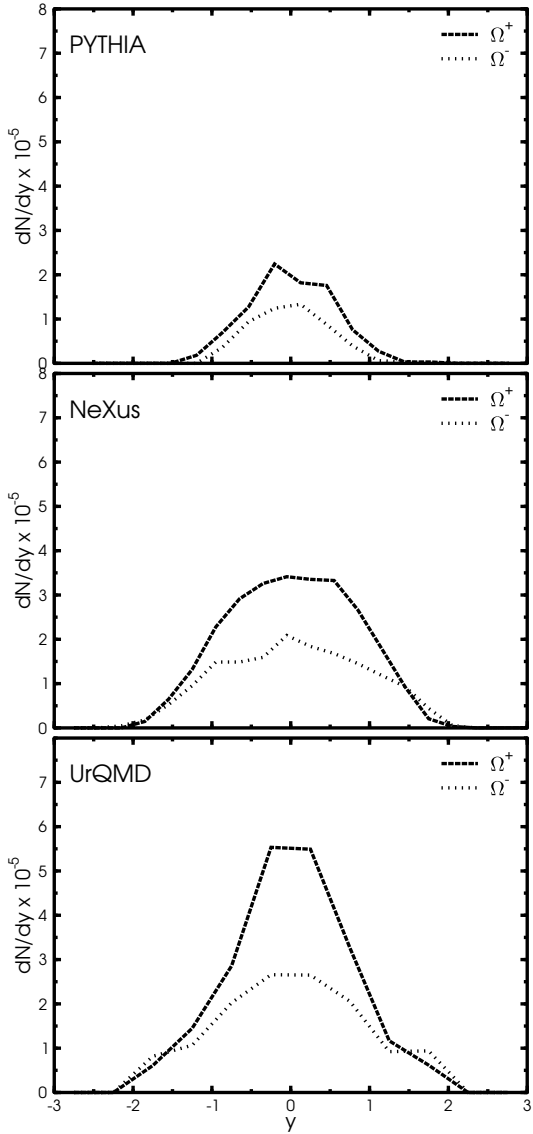,width=3.2in}} \vskip 2mm
\caption{Rapidity density of $\overline \Omega$ and $\Omega$ in pp
interactions at 160 GeV as predicted by UrQMD, NeXuS, PYTHIA.
\label{dndy}} \vspace{-0.4cm}
\end{figure}

The rapidity dependence of the $\Omega$ and $\overline{\Omega}$
yield is presented  in Fig.~\ref{dndy} within different string
models. The results were calculated in pp interactions at 160 GeV
within PYTHIA, NeXuS and UrQMD (from top to bottom). As can be
seen, the $\overline{\Omega}/\Omega$ ratio is largest around mid
rapidity.

The $\overline{\Omega}/\Omega$ ratio is fairly robust -- different
string-model implementations (PYTHIA, UrQMD, NeXuS) all agree in
their predictions within $\pm 20\%$. However, as shown in Table I
the total $4\pi$ yields of $\Omega^,$s and $\overline \Omega^,$s
may differ by a factor of 4 between the different string models.
The statistical models give in   general  more consistent results,
however, deviations up to $30\%$   are not excluded. In string
models, the particle abundances depend on the parameters chosen
for the fragmentation scheme, while in statistical models they
reflect the differences between the ensembles chosen. Thus, the
absolute yields allow a  distinction to be made  between the
implementations once experimental data become available. We will
show now that this is a generic feature that string models give
$\overline{\Omega}/\Omega>1$, whereas statistical models yield
$\overline{\Omega}/\Omega<1$ in pp interactions.


\begin{figure}[h]
\vskip 0mm \vspace{0cm}
\centerline{\psfig{figure=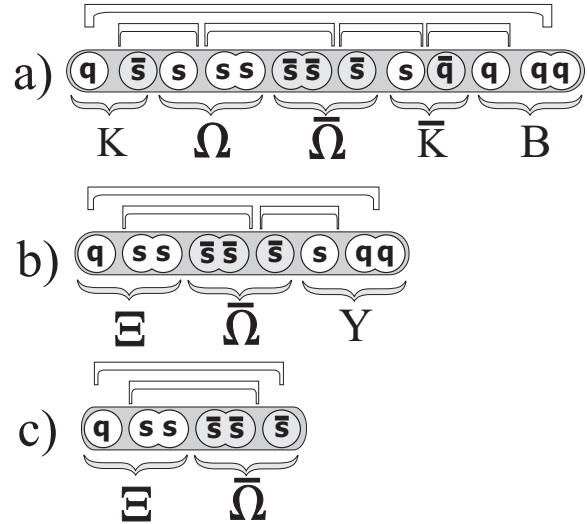,width=3in}} \vskip 2mm
\caption{Fragmentation of a color field into quarks and hadrons.
While in large strings $\Omega^,$s and $\overline \Omega^,$s are
produced in equal abundance (a), small di$^,$quark strings
suppress $\Omega^,$s at the string ends (b), sea-$\overline s$
quarks enhance $\overline \Omega^,$s (c).\label{string}}
\vspace{-0.4cm}
\end{figure}

In order to   understand  the large  $\overline{\Omega}/\Omega$
values predicted by string models one elucidates  in Fig.~
\ref{string} the color flux tube break-up mechanism. Fig.~
\ref{string} shows the fragmentation of the color field into
quark-antiquark pairs, which then coalesce into hadrons. While in
large strings $\Omega^,$s and $\overline \Omega^,$s are produced
in equal abundance (a), low-mass strings in UrQMD suppress
$\Omega$ production at the string ends (b), while in NeXuS
$\overline \Omega$$^,$s are enhanced (c). Thus, the microscopic
method of hadronization leads to a strong imbalance in
$\overline{\Omega}/\Omega$ ratio in low-mass strings.

The $\overline{\Omega}/\Omega$ ratio depends in a strongly
non-linear fashion on the mass of the fragmenting string. Fig.~
\ref{om} shows the $\overline \Omega/\Omega$ ratio as a function
of the mass of the fragmenting string (i.e. different beam
energies in pp). One clearly observes a strong enhancement of
$\overline {\Omega}$ production at low energies, while for large
string masses the ratio approaches the value  of
$\overline{\Omega}/\Omega = 1$ (which should be reached in the
limit of an infinitely long color flux tube).

Statistical models, on the other hand, are not able to yield a
ratio of $\overline{\Omega}/\Omega > 1$. This can be easily
understood in the GC formalism, where the $\overline{B}/B$ ratio
is very sensitive to the baryon chemical potential $\mu_B$. For
finite baryon densities, the $\overline{B}/B$ ratio will always be
$< 1$ and only in the limit of $\mu_B = 0$ may
$\overline{\Omega}/\Omega = 1$ be approached. These features
survive in the canonical framework, where the GC fugacities are
replaced by ratios of partition functions \cite{becattini97a,hag}.
This is shown in Fig.~\ref{om} (right) where the ratio
$\overline{\Omega}/\Omega$ in pp collisions (according to the
previously described model {\bf I}) is plotted as a function of
volume for four different temperatures. Hence, finite size
corrections in the statistical model actually lead to the opposite
behavior\cite{antik} in the ratio of $\overline{\Omega}/\Omega$
vs. system size (i.e. volume replacing string mass) to that
observed in the fragmenting color flux tube picture.
\begin{figure}[h]
\vskip 0mm
\centerline{\psfig{figure=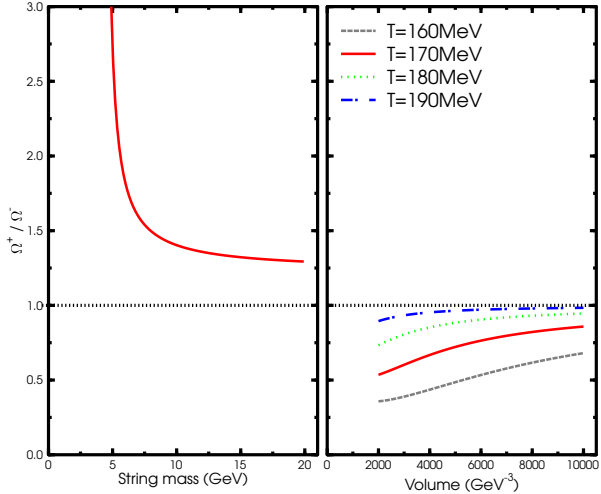,width=3.2in}} \vskip -3mm
\caption{Left: $\overline \Omega/\Omega$ ratio as a function of
string mass. Right: $\overline \Omega/\Omega$ ratio as a function
of the volume in model {\bf I}. \label{om}}
\end{figure}

In conclusion, within the fragmenting color flux tube models we
have predicted that the $\overline \Omega/\Omega$ ratio is
significantly above unity. This is in strong contrast to
statistical model results, which  always imply that $\overline B /
B $ ratios are below or equal to unity in proton proton reactions.
Since this  observable is accessible to NA49 measurements at the
SPS it can provide an excellent test to distinguish the
statistical model hadronization scenario from that of microscopic
color-flux tube dynamics.

We thank K. Kadjio (NA49) for fruitful and stimulating
discussions. S.B. acknowledges support from RIKEN, Brookhaven
National Laboratory and DOE grants DE-FG02-96ER40945 as well as
DE-AC02-98CH10886. K.R acknowledges Polish Committee for
Scientific Research grants  KBN-2P03B-03018. M.B. acknowledges
support from the region Pays de la Loire.

\end{document}